\documentclass{PoS}

\usepackage{braket}
\usepackage{subfigure}
\renewcommand{\d}{\ensuremath{\mathrm{d}}}

\title{Spectral densities from the lattice}

\ShortTitle{Spectral densities from the lattice}

\author{\speaker{Paulo J. Silva} \\
        Centro de F\'{i}sica Computacional, Departamento de F\'{i}sica, Universidade de Coimbra, 3004-516 Coimbra, Portugal\\
        E-mail: \email{psilva@teor.fis.uc.pt}}

\author{David Dudal \\
        Ghent University, Department of Physics and Astronomy, Krijgslaan 281-S9, 9000 Gent, Belgium\\
        E-mail: \email{david.dudal@ugent.be}}

\author{Orlando Oliveira \\
        Centro de F\'{i}sica Computacional, Departamento de F\'{i}sica, Universidade de Coimbra, 3004-516 Coimbra, Portugal\\
        E-mail: \email{orlando@fis.uc.pt}}

\abstract{We discuss a method to extract the K\"all\'{e}n-Lehmann spectral density of a particle (be it elementary or bound state) propagator by means of 4d lattice data. We employ a linear regularization strategy, commonly known as the Tikhonov method with Morozov discrepancy principle. An important virtue over the popular maximum entropy method is the possibility to also probe unphysical spectral densities, as, for example, of a confined gluon. We apply our proposal to the SU(3) glue sector.}

\FullConference{31st International Symposium on Lattice Field Theory LATTICE 2013\\
		 July 29 - August 3, 2013\\
		 Mainz, Germany}

\begin{document}

\section{Introduction}

In recent years, there has been an renewed interest in the infrared behaviour
of the Landau gauge Yang-Mills propagators, related to the gluon confinement
phenomenon. In particular, lattice studies of the Landau gauge gluon propagator
\begin{equation}
D^{ab}_{\mu\nu} ( \hat{q} ) ~ = ~ \delta^{ab} ~
   \Big( \delta_{\mu\nu} ~ - ~ \frac{q_\mu q_\nu}{q^2} \Big) ~
   D( q^2 ) ~ ,\label{propcont}
\end{equation}
have been performed at large volumes, with the propagators reaching a finite
non-zero value in the infrared region. While the simulations have been
performed using volumes as large as $(27~\textrm{fm})^4$ for the SU(2) gauge
group \cite{cucc07} and $(17~\textrm{fm})^4$ for the SU(3) gauge group \cite{bma09},
the lattice spacing used in the simulations was quite big,
being $0.22$ fm for SU(2) and $0.18$ fm for SU(3).
In a recent paper \cite{olisi12} by some of us, numerical evidence
has been given that a large lattice spacing also changes the propagator in the infrared region.

Gluons are not physical particles. Besides the mere question of the infrared behaviour of their propagator, one would also like to identify signs of gluon confinement in this two-point function.

It is already known, from lattice simulations \cite{cucc05, aubin04, sioli06, bowman07},
that the Landau gauge gluon propagator displays a violation of
spectral positivity.
This implies that the gluon cannot appear as a free asymptotic S-matrix
state and may be viewed as an indication of gluon confinement.
A mini-review about gluon positivity violation can be found in \cite{cornwall}.

Lattice studies of positivity proceed by studying the Schwinger function
\begin{equation}
  C(t) = \int_{-\infty}^{\infty} \frac{\d p}{2\pi} D(p^2) \exp(-ipt).
\label{schwinger}
\end{equation}
Assuming a K\"all\'en-Lehmann spectral representation for the gluon, we get
\begin{equation}
  C(t)=  \int_{0}^{\infty} \d\omega \rho(\omega^2) e^{-\omega t},
\label{kallen}
\end{equation} 
a quantity shown in Figure 1. Note that, although $C(t) < 0$ implies a negative spectral density
$\rho(\omega^2)$, and hence positivity violation, a positive $C(t)$
says nothing about the sign of the spectral density.

\begin{figure}[t]
\begin{center}
\vspace*{0.5cm}
\includegraphics[width=0.5\textwidth]{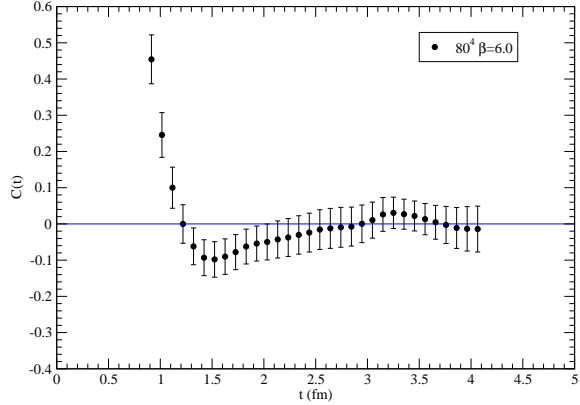}
\caption{Temporal correlator for the gluon propagator computed using $80^4$ $\beta=6.0$ lattice data.}
\end{center}
\end{figure}

\section{Gluon spectral densities}

Generally speaking, if $\mathcal{G}(p^2)\equiv\braket{\mathcal{O}(p)\mathcal{O}(-p)}$ is an
 Euclidean momentum-space propagator of a (scalar) physical degree of freedom, then it must have a K\"{a}ll\'{e}n-Lehmann spectral representation
\begin{equation}
\mathcal{G}(p^2)=\int_{0}^{\infty}\d\mu\frac{\rho(\mu)}{p^2+\mu}
\,,%\qquad \text{with }\rho(\mu)\geq0 \text{ for } \mu\geq 0\,.
\label{specdens}
\end{equation}
with $\rho(\mu)\geq0$ for $\mu\geq 0\,$. The spectral density contains
information on the masses of physical states described by the
operator $\mathcal{O}$.

Our goal is to compute the spectral density $\rho(\mu)$ from the
propagator $\mathcal{G}(p^2)$. We can express eq.~(\ref{specdens})
as a double Laplace transform
$  \mathcal{G}=\mathcal{L}^2\hat\rho=\mathcal{L}\mathcal{L}^\ast \hat\rho$
where $(\mathcal{L}f)(t)\equiv\int_0^{\infty}\d s e^{-st}f(s)$.
This is a notorious ill-posed problem. Note that for
$\mathcal{G}(p^2)$ one usually has a set of data points with error bars.

A popular approach to compute spectral functions has been the
maximum entropy method \cite{mem}. Here we use an alternative approach,
based on Tikhonov regularization with Morozov discrepancy principle.
For details see our previous papers \cite{latt2012, qchsx, letter}.

To compute the spectral density we consider the integral equation
\begin{equation}
    \int_0^\infty\d t\rho(t)\frac{\ln\frac{z}{t}}{z-t}+\lambda\rho(z)=\int_0^\infty \d t \frac{\mathcal{D}(t)}{t+z}
\label{system-old}
\end{equation}
where for $\mathcal{D}(t)$ we use lattice data in momentum space for the
gluon propagator computed in a $80^4$ volume, with $\beta=6.0$ \cite{olisi12}.
In a loose way of speaking, we search for a solution to the integral
eq.~(\ref{specdens}) that is sufficiently close in norm to the exact but
unknown spectral function. This ``sufficiently close'' translates
into a variational problem, which solution fulfills the foregoing
eq.~(\ref{system-old}).  The r\^{o}le of the parameter $\lambda>0$ is 
to regularize the ill-posed nature of the problem. 
For practical purposes, the finite set of data points was interpolated using
splines. Furthermore, one has to impose IR and UV cut-offs, although we
consider a 1-loop perturbative behaviour after $p^{(latt)}_{max}$.
The integrals in eq.~(\ref{system-old}) are computed using Gauss-Legendre
quadrature, and the discretization of eq.~(\ref{system-old}) leads to a
linear system one has to solve in order to compute the spectral density.

\begin{figure}[h]
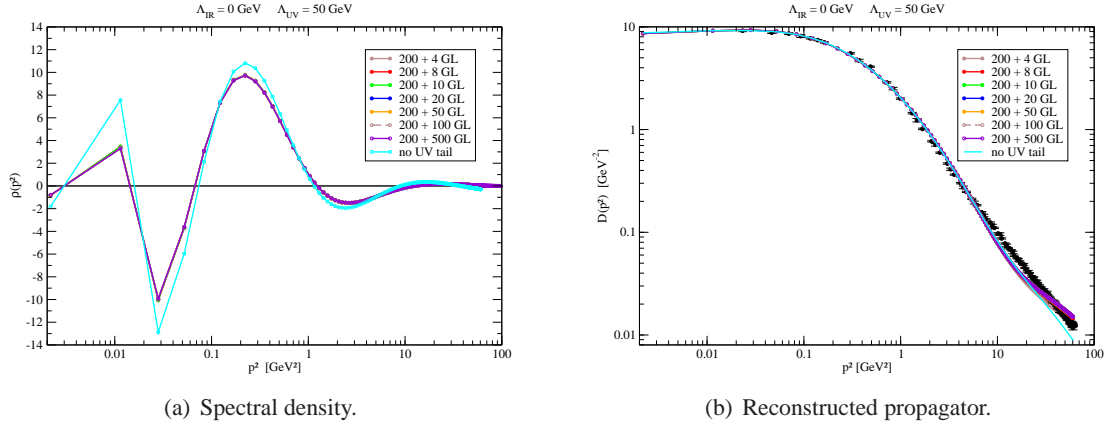
 %  figure placement: here, top, bottom, or page   (scale was 0.235)
   \centering
   \subfigure[Spectral density.]{ \includegraphics[scale=0.28]{plots/spectral/Tzero/rho_200_various} } \qquad
   \subfigure[Reconstructed propagator.]{ \includegraphics[scale=0.28]{plots/spectral/Tzero/prop_200_various} }
  \caption{Results for the spectral density and for the reconstructed propagator with a different number of Gauss-Legendre points.}
   \label{specT0-1}
\end{figure}

\begin{figure}[h]
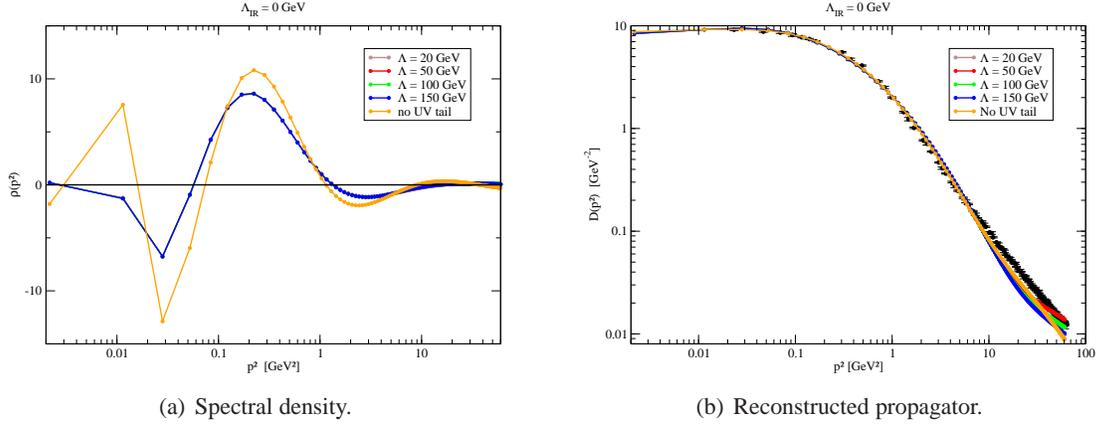
 %  figure placement: here, top, bottom, or page
\vspace{0.2cm}
   \centering
   \subfigure[Spectral density.]{ \includegraphics[scale=0.28]{plots/spectral/Tzero/rho_200_10} } \qquad
   \subfigure[Reconstructed propagator.]{ \includegraphics[scale=0.28]{plots/spectral/Tzero/prop_200_10} }
  \caption{The effect of changing the UV cutoff.}
   \label{specT0-2}
\end{figure}

\begin{figure}[h]
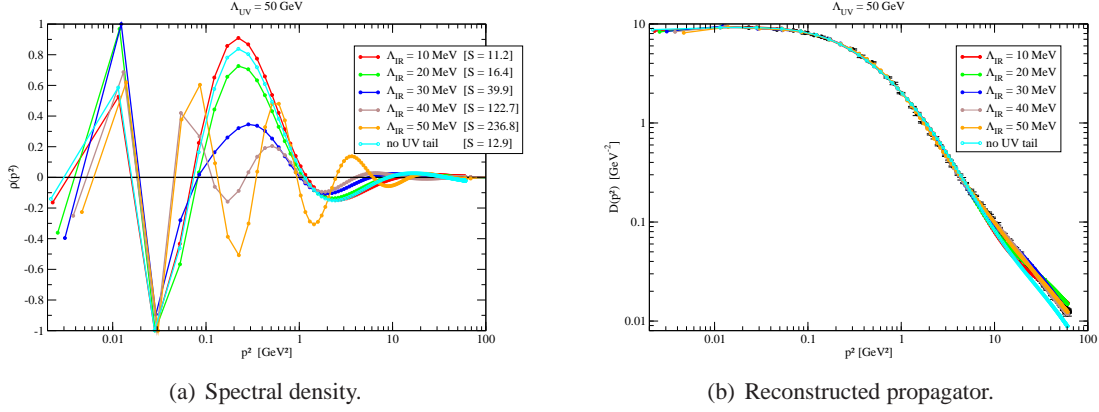
 %  figure placement: here, top, bottom, or page
   \centering
   \subfigure[Spectral density.]{ \includegraphics[scale=0.27]{plots/spectral/Tzero/rho_200_20_LIR} } \qquad
   \subfigure[Reconstructed propagator.]{ \includegraphics[scale=0.27]{plots/spectral/Tzero/prop_200_20_LIR} }
  \caption{The effect of changing the IR cutoff.}
   \label{specT0-3}
\end{figure}

In Figures \ref{specT0-1}, \ref{specT0-2},  and \ref{specT0-3} we study how the solution to
eq.~(\ref{system-old}) depends on the number of Gauss-Legendre (GL) points,
and on the IR and UV cut-offs. In what concerns the number of GL points and
the UV cut-off, the results for the spectral density (left-hand plots)
show that they do not have a big influence in
the results. However, we see some differences in the results if we do not
consider a UV tail beyond the maximum momentum available in the lattice
simulation considered here. Nevertheless, the reconstruction of the input
propagator (right-hand plots) apparently does not distinguish the several
spectral functions.

On the other hand, a change in the IR cut-off has a dramatic effect in the
computed spectral function. Note that in Figure \ref{specT0-3}, the 'S' values
refer to the scale factors needed to fit the spectral density in the interval
$[-1,1]$. Despite this, the reconstructed propagators are still
indistinguishable.

\begin{figure}[h] %  figure placement: here, top, bottom, or page
   \centering
   \subfigure[Spectral density.]{ \includegraphics[scale=0.26]{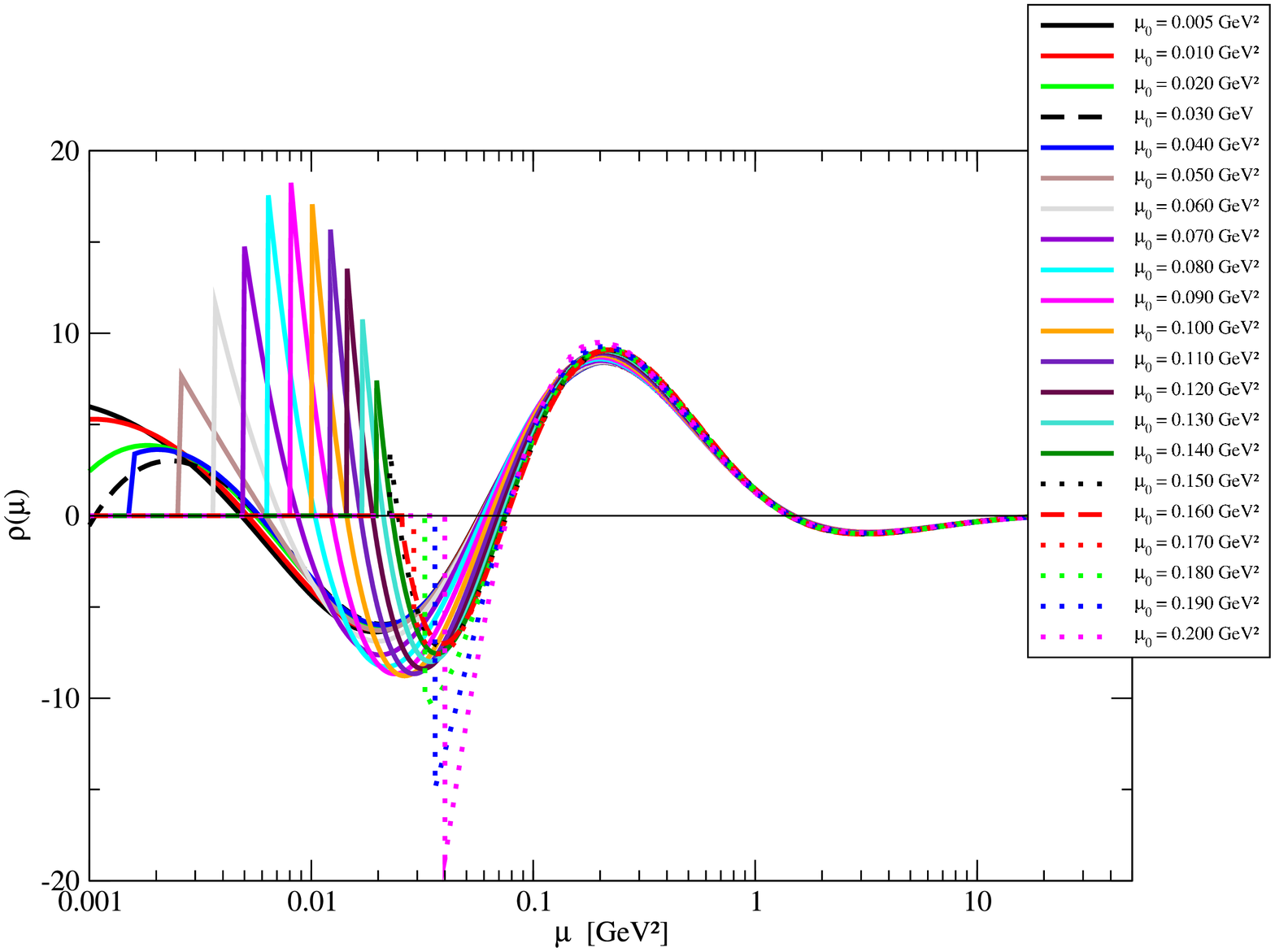} } \qquad
   \subfigure[Reconstructed propagator.]{ \includegraphics[scale=0.26]{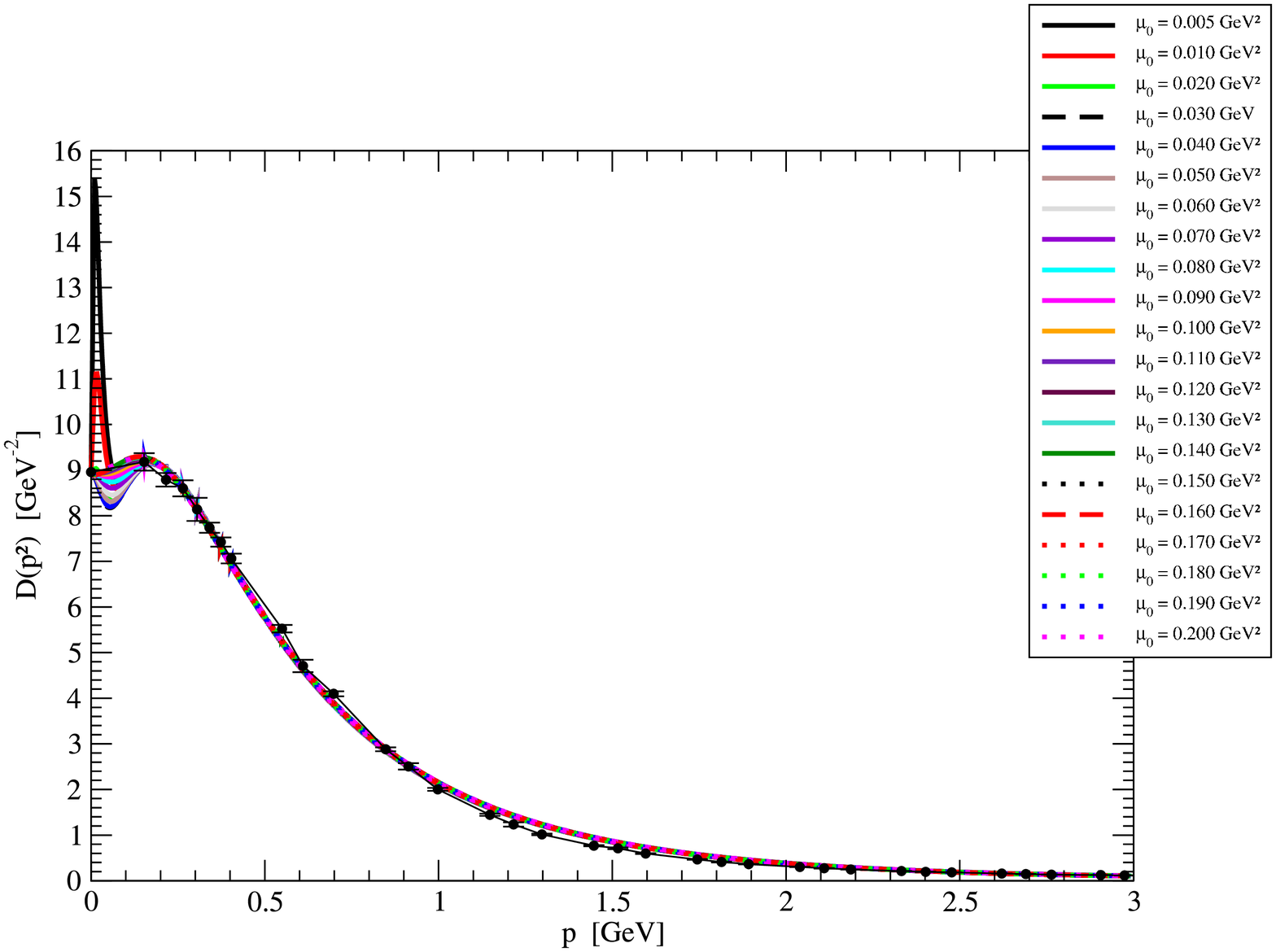} }
  \caption{Results obtained using the updated method presented in \cite{letter}.}
   \label{specT0-new}
\end{figure}

This study leaded us to try to find out a clever way to compute the spectral
density, and hopefully an updated method arose -- see \cite{letter} for
details. Although there are natural differences in the deep infrared region
due to the different choices of the IR cut-off $\mu_0$, the results are much
more stable if compared with previous results -- see  Figure \ref{specT0-new}.
Moreover, the IR cut-off is now interrelated with the
Tikhonov parameter $\lambda$ -- see \cite{letter} for details.

Given recent results for the gluon spectral function obtained from
the solution of the Dyson-Schwinger equations in the complex momentum
plane \cite{fischer2012}, a comparison with our results is in order.
It turns out that there are fundamental differences between the results
of the two works. In particular, we do not see evidence for the
sharp peak reported in \cite{fischer2012}. Furthermore, our results
show a violation of positivity setting in for small momenta, whereas
in \cite{fischer2012} positivity violation occurs only for momenta
above 600 MeV.

\section{Results at finite temperature}

In this section, we consider lattice results for the gluon propagator at
finite temperature, and study positivity violation through the computation
of the Schwinger and spectral functions.

The lattice setup for the simulations at finite temperature considered
here is reported in \cite{gluonmass}.

\subsection{Positivity violation}

\begin{figure}[h]
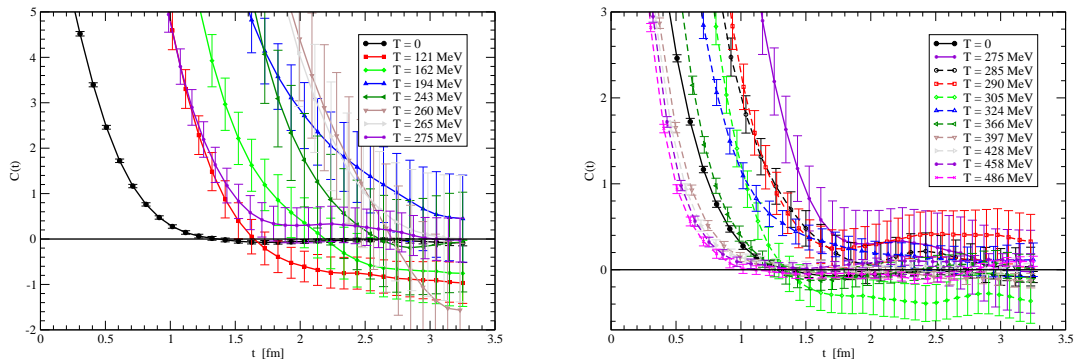
 %  figure placement: here, top, bottom, or page
\vspace*{0.3cm}
   \centering
   \subfigure{ \includegraphics[scale=0.28]{plots/spectral/Tfinita/positividade_long_upTc} } \qquad
   \subfigure{ \includegraphics[scale=0.28]{plots/spectral/Tfinita/positividade_long_aboveTc} }
  \caption{Temporal correlator for the longitudinal component.}
   \label{poslong}
\end{figure}

\begin{figure}[h]
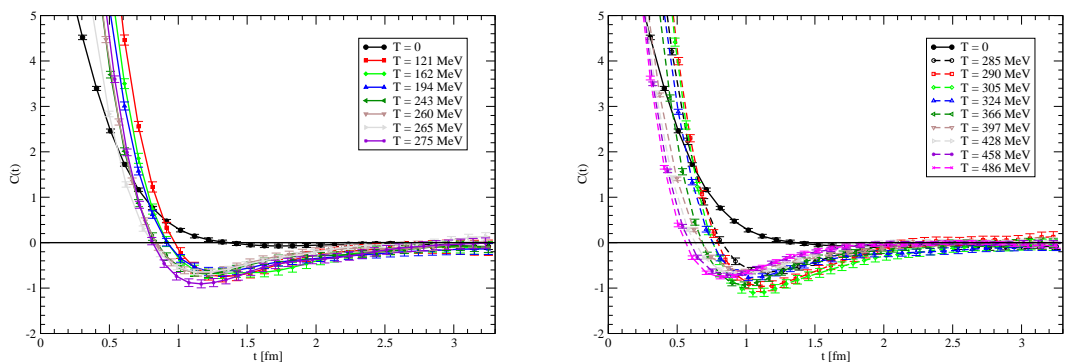
 %  figure placement: here, top, bottom, or page
\vspace*{0.6cm}
   \centering
   \subfigure{ \includegraphics[scale=0.28]{plots/spectral/Tfinita/positividade_trans_upTc} } \qquad
   \subfigure{ \includegraphics[scale=0.28]{plots/spectral/Tfinita/positividade_trans_aboveTc} }
  \caption{Temporal correlator for the transverse component.}
   \label{postrans}
\end{figure}

\begin{figure}[h]
\vspace*{0.3cm}
\begin{center}
\includegraphics[width=0.5\textwidth]{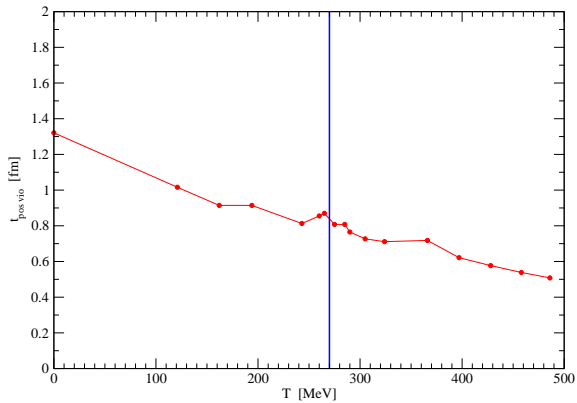}
\caption{Positivity violation scale for the transverse propagator.}
\label{viotrans}
\end{center}
\end{figure}

In this subsection, the temporal correlator defined in eq.~(\ref{schwinger}) is
computed for the longitudinal and transverse components of the gluon propagator
for all available temperatures. From the results shown in
Figures \ref{poslong} and \ref{postrans}, it turns out that positivity is
violated for both transverse and longitudinal components at all temperatures.
Note that, in what concerns the transverse propagator, the time scale for
positivity violation decreases with the temperature -- see Figure \ref{viotrans}. This suggests that, for sufficiently high
temperatures, transverse gluons can behave as quasi-particles.

\begin{figure}[t]
\vspace*{0.5cm}
\begin{center}
\includegraphics[width=0.6\textwidth]{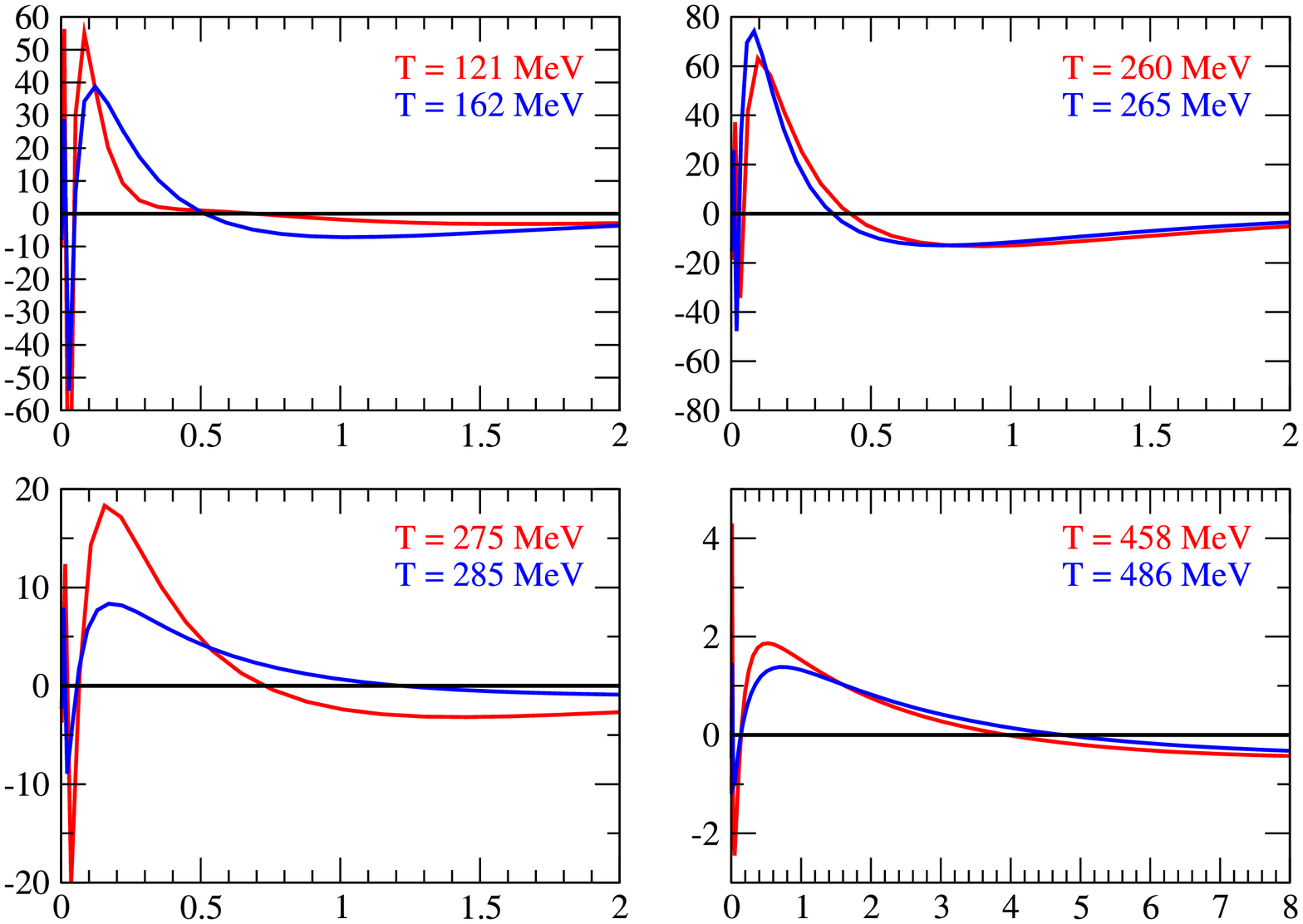}
\caption{Longitudinal propagator spectral densities.}
\label{specT}
\end{center}
\end{figure}

\subsection{Spectral densities}

In this subsection, we consider the spectral densities, computed from eq.~(\ref{system-old}), for the longitudinal component -- see
Figure \ref{specT}.  Notice that the energy scale at which the
axis is crossed, increases with temperature. This means, again, that
for sufficiently high temperatures, longitudinal gluons may be
considered quasi-particles.

\section*{Acknowledgments}
O.~Oliveira and P.~J.~Silva acknowledge support by FCT via projects
CERN/FP/123612/2011, CERN/FP/123620/2011, and PTDC/FIS/100968/2008,
developed under the initiative QREN financed by the UE/FEDER through
the Programme COMPETE - Programa Operacional Factores de Competitividade.
P.~J.~Silva is also supported by FCT grant SFRH/BPD/40998/2007.
D.~Dudal acknowledges financial support from the Research-Foundation
Flanders (FWO Vlaanderen) via the Odysseus grant of F.~Verstraete.

\end{document}